\documentclass[pra,showpacs,superscriptaddress,twocolumn]{revtex4-1}
\usepackage[dvipdfmx]{graphicx}%RH付け足し 160115arXiv用に消した　160117復活　160517arxiv-v3のため消した 180622復活
\usepackage{amsmath}
 
\graphicspath{{graphics/}}  %%RH付け足し  160115arXiv用に消した　160117バリでコンパイルするために復活 160517arxiv-v3のため消した 180622復活
\usepackage{color}
\usepackage{cancel}
\usepackage{ulem}
\usepackage{comment} %%RH180622追加
\usepackage{bm} %%RH190225追加
\usepackage{braket}

\newcommand{\MHz}[1]{{#1}{\,\rm{MHz}}}

\newcommand{\micron}[1]{{#1}{\,\rm{\mu m}}}

\newcommand{\dBm}[1]{{#1}{\,\rm{dBm}}}

\newcommand{\LiNb}{\rm{LiNbO}_3}

\usepackage{float}

\begin{document}
\title{Quantitative optical imaging method for surface acoustic waves \\using optical path modulation}

\author{Ryusuke~Hisatomi}
\email{hisatomi.ryusuke.2a@kyoto-u.ac.jp}
\affiliation{Institute for Chemical Research (ICR), Kyoto University, Gokasho, Uji, Kyoto 611-0011, Japan}
\affiliation{Center for Spintronics Research Network (CSRN), Kyoto University, Gokasho, Uji, Kyoto 611-0011, Japan}
\affiliation{PRESTO, Japan Science and Technology Agency, Kawaguchi-shi, Saitama 332-0012, Japan}
\author{Kotaro~Taga}
\affiliation{Institute for Chemical Research (ICR), Kyoto University, Gokasho, Uji, Kyoto 611-0011, Japan}
\author{Ryo~Sasaki}
\affiliation{RIKEN Center for Quantum Computing (RQC), RIKEN, Wako, Saitama 351-0198, Japan}
\author{Yoichi~Shiota}
\affiliation{Institute for Chemical Research (ICR), Kyoto University, Gokasho, Uji, Kyoto 611-0011, Japan}
\affiliation{Center for Spintronics Research Network (CSRN), Kyoto University, Gokasho, Uji, Kyoto 611-0011, Japan}
\author{Takahiro~Moriyama}
\affiliation{Institute for Chemical Research (ICR), Kyoto University, Gokasho, Uji, Kyoto 611-0011, Japan}
\affiliation{Center for Spintronics Research Network (CSRN), Kyoto University, Gokasho, Uji, Kyoto 611-0011, Japan}
\affiliation{PRESTO, Japan Science and Technology Agency, Kawaguchi-shi, Saitama 332-0012, Japan}
\author{Teruo~Ono}
\affiliation{Institute for Chemical Research (ICR), Kyoto University, Gokasho, Uji, Kyoto 611-0011, Japan}
\affiliation{Center for Spintronics Research Network (CSRN), Kyoto University, Gokasho, Uji, Kyoto 611-0011, Japan}

\date{\today}

\begin{abstract}
A Rayleigh-type surface acoustic wave (SAW) is used in various fields as classical and quantum information carriers because of its surface localization, high electrical controllability, and low propagation loss.
Coupling and hybridization between the SAW and other physical systems such as magnetization, electron charge, and electron spin are the recent focuses in phononics and spintronics.
A precise measurement of the surface wave amplitude is often necessary to discuss the coupling strengths. However, there are only a few such measurement techniques and they generally require a rather complex analysis.
Here we develop and demonstrate a straightforward measurement technique that can quantitatively characterize the SAW. The technique optically detects the surface waving due to the coherently driven SAW by the optical path modulation. Furthermore, when the measurement system operates in the shot-noise-limited regime, the surface slope and displacement at the optical spot can be deduced from the optical path modulation signal.
Our demonstrated technique will be an important tool for SAW-related research.
%221116
\end{abstract}


\maketitle

\section{Introduction}
Rayleigh-type surface acoustic wave (SAW) is a propagating wave on a surface of elastic material~\cite{LL7,TB2017}.
Due to their surface localization, high electrical controllability, and low propagation loss, the SAW is used in many classical electrical devices, e.g., communication equipment and sensors for gases and liquids~\cite{C1998,BB2008,LR2008,RM2009,R2017,ZL2021}.
The SAW is also a quantum information carrier, taking advantage of its long coherence time~\cite{D2019,SK2015,MK2017,MS2018,SZ2018,NY2020}.
Recently, the coupling and hybridization of the SAW with other physical systems also attracted attention, including magnetoelastic coupling~\cite{W2002,WD2011,WH2012,GM2015,BV2020} between the SAW and magnetization, acoustoelectric coupling~\cite{MC2012,PB2015,PS2015} between the SAW and electron charge, and spin-vorticity coupling~\cite{M2013,IMM2014,K2017,K2020} between the SAW and electron spin.
In addition, the phenomena that emerged from the angular momenta associated with the SAW also attracted attention~\cite{LR2018,SK2021,B2022}.
%221116

Quantitative imaging of the SAW is a powerful tool for advancing SAW-related research.
Specifically, it leads to the evaluation of various coupling strengths as well as the characterization of the SAW itself.
A few optical measurement techniques were developed to realize this~\cite{K2000,H2011,FS2019,K2005,H1970,FS2019,IN2022}, but it is challenging to obtain quantitative results in practice in many cases.
Thus, we turned our attention to a simple phenomenon in which an optical path of reflected light is modulated according to the law of reflection depending on a surface slope at the optical spot position.
We developed a method to quantitatively obtain spatial distributions of the amplitude and phase of the surface slope by observing this optical path modulation with a measurement system working in the shot-noise-limited regime.
Those distributions provide enough information to specify the SAW with a wavelength longer than the optical spot size.
Although our recently reported imaging method of the SAW using a polarimeter can also quantitatively evaluate the surface slope~\cite{TH2021}, it requires the material complex refractive index at the optical spot position for analysis.
That might be a problem in experiments with novel materials.
The method using the optical path modulation is, therefore, more versatile.
%221116

This paper shows that the optical path modulation method can quantitatively evaluate the amplitude and phase of a time-periodically varying surface slope due to the coherently driven SAW.
Furthermore, by sweeping the optical spot within the device surface, two-dimensional spatial imaging of the amplitude and phase of the time-varying surface slope is realized.
%221116

After a brief description of a theoretical model of the quantitative probing of the surface slope in Sec.~\ref{sec:model}, we present experiments in which quantitative measurement of the surface slope with the SAW is demonstrated in Sec.~\ref{sec:experiments}.
Then, the slope amplitude obtained from the optical path modulation signal is compared with the slope amplitude obtained from other schemes, and its validity is discussed.
Finally, the surface displacement is estimated from the spatial distribution of the slope amplitude.
%221116

%%%% Figure 1%%%%
\begin{figure*}[t]
\begin{center}
\includegraphics[width=17.8cm,angle=0]{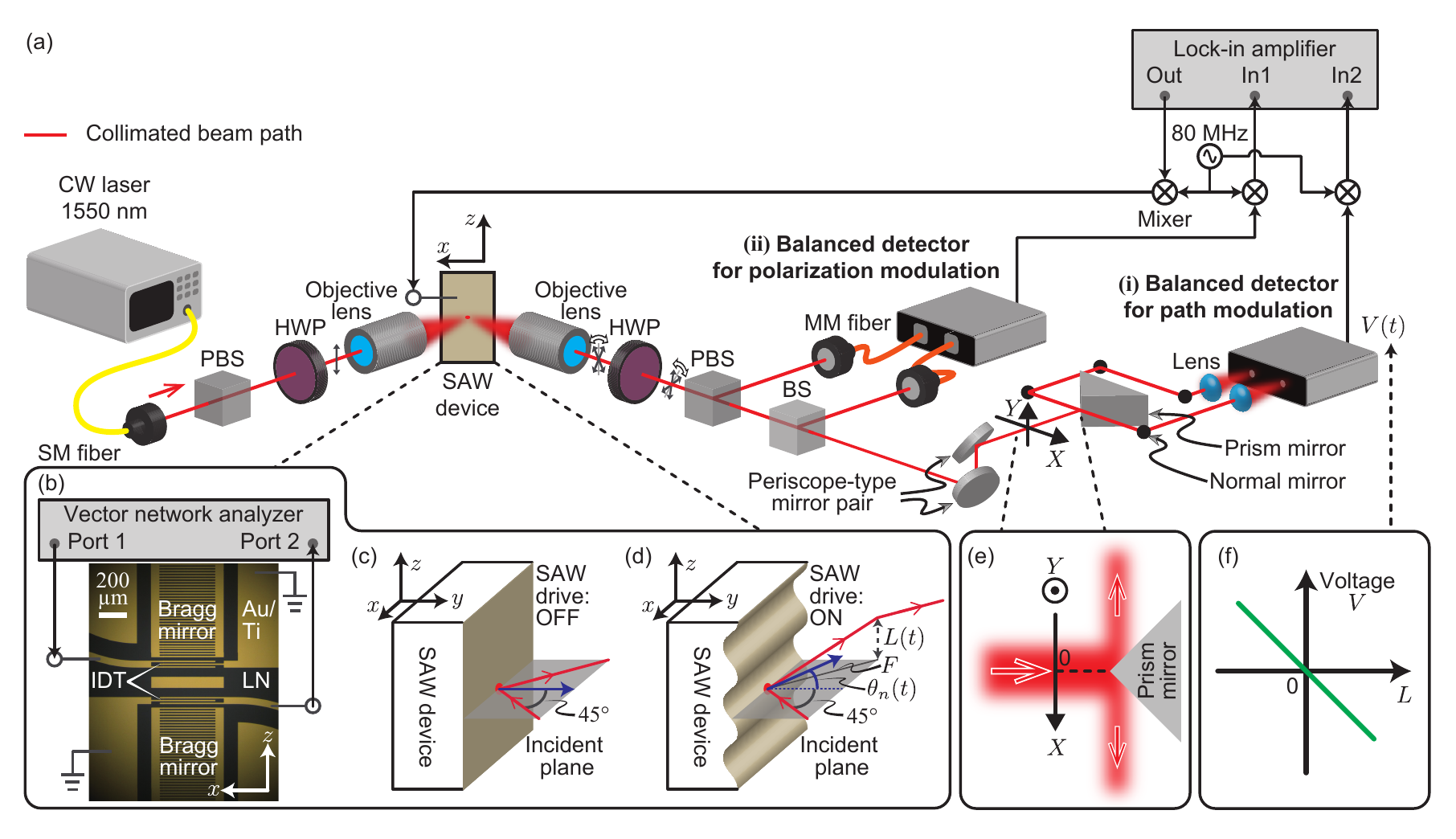}
\caption{
Experimental setup and device for observing optical path modulation and optical polarization modulation induced by SAW. (a) With polarization beam splitter (PBS) and half-wave plate (HWP), the incident light is made linearly polarized with the polarization plane aligned parallel to the direction of SAW propagation
($z$-direction). The light is focused with an objective lens into a diameter of around $3\,\rm{\mu}m$ on the device surface. The angle of incidence is set to $45^{\circ}$. The reflected light is then collimated by another objective lens and analyzed in the two balanced detectors indicated as (i) and (ii) with HWP, PBS, beam splitter (BS), periscope-type mirror pair, and prism mirror. 
(b) Microscope image of a Fabry-Perot type SAW resonator with a coordinate system. The SAW resonator consists of a $0.5$-mm-thick YZ-cut $\LiNb$ substrate, Bragg mirrors, and interdigitated transducers (IDTs).
Bragg mirrors and IDTs are made by depositing $5\,\mathrm{nm}$ of Ti and $80\,\mathrm{nm}$ of Au. Two Bragg mirrors composed of $100$ fingers of $400$-$\mathrm{\mu m}$-long with a $10$-$\mathrm{\mu m}$ line and space are placed at a distance of $360\,\mathrm{\mu m}$. 
Two IDTs composed of three fingers of $400$-$\mathrm{\mu m}$-long with a $10$-$\mathrm{\mu m}$ line and space are placed $105\,\mathrm{\mu m}$ apart from the center between Bragg mirrors, respectively.
A Ti($5\,\mathrm{nm}$)/Au($80\,\mathrm{nm}$) film with a length of $520\,\mathrm{\mu m}$ and a width of $80\,\mathrm{\mu m}$ is deposited at the center between Bragg mirrors. The SAW resonator is electrically characterized by rf transmission measurement from Port 1 to Port 2 of a vector network analyzer through two IDTs.
(c,d) Schematics of light reflection on the surface of the SAW device when the SAW drive is (c) OFF and (d) ON.
(e) Setup for measuring beam displacement in the $X$-direction induced by optical path modulation by the SAW. (f) Relationship between slight beam displacement $L$ and output voltage $V$ from the balanced detector.
%221116
}
\label{fig:setup}
\end{center}
\end{figure*}

\newpage
\section{Measurement principle of \\surface slope} \label{sec:model}

According to the law of reflection, a time variation in the local slope of the device surface induces a time modulation of the light path reflected off that local position.
This phenomenon, \textit{optical path modulation}, is the key to surface slope measurement.
Figure~1 summarizes the schematics of this paper's experimental setup and diagrams related to the measurement principle.
Figures~1(c) and 1(d) show the experimental situation around the SAW device.
The red arrows represent wave vectors of the incident and reflected light, and the blue arrows represent normal vectors at the optical spot.
Although the optical path modulation occurs at any incident angle, we set the incident angle to $45^{\circ}$ and the propagation direction of the SAW in the $z$-direction.
In this situation, we can observe the optical path and the polarization modulation simultaneously. 
By comparing the slope amplitudes obtained from both modulation measurements, the validity of the optical path modulation measurement can be verified. 
%221116

From here, we describe the details of the optical path modulation measurement according to the experimental situation in Fig.~\ref{fig:setup}(a).
Let us assume that the normal vector of the SAW device surface in Fig.~\ref{fig:setup}(c) is initially oriented in the $y$-direction, and the wave vector of the incident light is in the $xy$-plane.
In this case, the wave vector of the reflected light is definitely in the $xy$-plane.
On the other hand, when we drive the SAW propagating in the $z$-direction shown in Fig.~\ref{fig:setup}(d), the normal vector is periodically tilted with time in the $z$-direction.
Accordingly, the wave vector of the reflected light deviates from the $xy$-plane, resulting in the optical path modulation.
Note here that the optical spot size needs to be smaller than the quarter wavelength of the SAW to allow the law of reflection to hold locally.
This paper achieves the condition by inserting an objective lens in front of the SAW device and placing the device surface at the focus, as shown in Fig.~1(a).
%221117

The tilt of the normal vector in Fig.~1(d) is the same as the surface slope, which can be written as
\begin{equation}
    \theta_n(x,z,t) = \frac{\partial u_{y}(x,z,t)}{\partial z},\label{eq:Tanduy}
\end{equation}
where $u_y(x,z,t)$ is the $y$-component of the surface displacement at the optical spot position $x$ and $z$ and time $t$.
As shown in Figs.~1(a) and~1(d), by placing an objective lens after the SAW device in the proper position (such that the reflected light is collimated), the wavevector of the reflected light becomes parallel to the $xy$-plane again.
In this case, the vertical displacement of the optical beam from the nominal position, $L(x,z,t)$, is given by
\begin{equation}
L(x,z,t) =F\times \tan\lbrack 2\theta_n(x,z,t)\rbrack \simeq F\times2\theta_n(x,z,t),\label{eq:LandTheta}
\end{equation}
where $F$ is the focal length of the objective lens after the SAW device and $\theta_{n}(x,z,t)$ is assumed to be very small.
%221117

%when a particular SAW mode is driven by electrical signal at angular frequency $\omega_{\rm{SAW}}$ 
To observe the surface slope $\theta_n(x,z,t)$, we measure the optical beam displacement $L(x,z,t)$ in the following manner by using a periscope-type mirror pair and subsequent elements in Fig~1(a).
First, the periscope-type mirror pair plays the functional role of converting vertical beam displacement along the $z$-axis into horizontal beam displacement along the $X$-axis before a prism mirror.
Second, as shown in Fig.~1(e), the prism mirror placed in the center of the beam splits the light in half without SAW driving.
Third, the split lights are collected on two photodetectors constituting a balanced detector using lenses.
Then, the differential voltage signal $V(x,z,t)$ output from the balanced detector is input to a lock-in amplifier.
%Under this situation, as shown in Fig.1(f), 
The output differential voltage signal $V(x,z,t)$, which is proportional to the difference in input optical power, can be written using the beam displacement $L(x,z,t)$ as follows,
\begin{equation}
V(x,z,t) =  2CDL(x,z,t),\label{eq:VandL}
\end{equation}
where $C$ is a conversion efficiency of the balanced detector and $D$ is an optical power density in the $X$-direction before the prism mirror.
Substituting Eq.~(\ref{eq:LandTheta}) into Eq.~(\ref{eq:VandL}), we find $V(x,z,t)=4CDF\theta_{n}(x,z,t)$.
Finally, by measuring the voltage signal $T\times V(x,z,t)$ with the lock-in amplifier, where $T$ is the transfer coefficient including attenuation and amplification of the electrical signal in the transmission line between the balanced detector and the lock-in amplifier, we can obtain information related to the surface slope $\theta_n(x,z,t)$. 
%221117

To evaluate the absolute value of the slope amplitude from the experiment, it is usually necessary to calibrate the measurement system.
Specifically, the values $T$, $C$, and $D$, for which no nominal values exist, need to be determined from independent experiments.
The procedure is usually rather difficult.
However, we show here that it is possible to evaluate the slope amplitude straightforwardly with our scheme as long as we can observe the \textit{optical shot noise}. The surface slope $\theta_n(x,z,t)$ is written as
\begin{equation}
\theta_n(x,z,t) =\mathrm{Re}\left[\theta_n(x,z) e^{-i\omega_{\rm{SAW}} t}\right], \label{eq:theta_xzt}
\end{equation}
where
\begin{equation}
  \left\{ 
  \begin{alignedat}{2}   
    \theta_{n}(x,z)&=&|\tilde\theta_{n}(x,z)|\,e^{i\varphi(x,z)}\,\,\, (\hbox{if propagating wave}), \\   
    \theta_{n}(x,z)&=&|\tilde\theta_{n}(x,z)|\,\sin\lbrack\varphi(x,z)\rbrack\,\,\, (\hbox{if F.-P. resonator}),
  \end{alignedat} 
  \right. \label{eq:theta_xz}
\end{equation}
is the position-dependent amplitude of $\theta_{n}(x,z,t)$ of a propagating wave or a wave in a Fabry-Perot-type SAW resonator. Note that $|\tilde\theta_{n}(x,z)|$ can be considered to be more or less constant and the spatial variation of $\theta_{n}(x,z)$ is encoded in the phase part $\varphi(x,z)$. Note also that $\omega_{\mathrm{SAW}}$ is an angular frequency of the SAW.
The key to our method is that the theoretical expression for the power spectrum acquired by the lock-in amplifier with SAW driven is given by (see Appendix \ref{SM:PS} for details),
\begin{equation}
    S_{VV}(x,z,\omega_{\mathrm{SAW}}) \Delta f \propto \frac{8D^2F^2|\theta_{n}(x,z)|^2}{\hbar^2\omega_{0}^2} +\frac{P_{i}}{\hbar \omega_0}\Delta f,\label{eq:PS}
\end{equation}
where $S_{VV}(x,z,\omega)$ is the power spectral density obtained when the optical spot is at position $(x,z)$, $\Delta f$ is a measurement bandwidth, $P_{i}$ is a power of the incident light going to the prism mirror, $\omega_0$ is an angular frequency of the incident light, and $\hbar$ is Planck's constant.
The first term represents the coherent signal due to the driven SAW, and the second term represents the optical shot noise.
In a situation where optical shot noise can be measured, we can experimentally obtain a ratio between the first and second terms in Eq.~(\ref{eq:PS}), $R$, defined by
\begin{equation}
    R(x,z)=\frac{8 D^2 F^2 |\theta_n(x,z)|^2}{P_{i} \hbar \omega_0 \Delta f }.\label{eq:R}
\end{equation}
The experimentally acquired ratio $R$ allows for a quantitative evaluation of the slope amplitude $|\theta_n(x,z)|$ with the known values of the focal length of the objective lens $F$, the optical power density $D$ (see Appendix \ref{SM:D} for details), the incident optical power $P_{i}$, and the measurement bandwidth $\Delta f$.
%221117

%%%% Figure 2%%%%
\begin{figure*}[t]
\begin{center}
\includegraphics[width=18cm,angle=0]{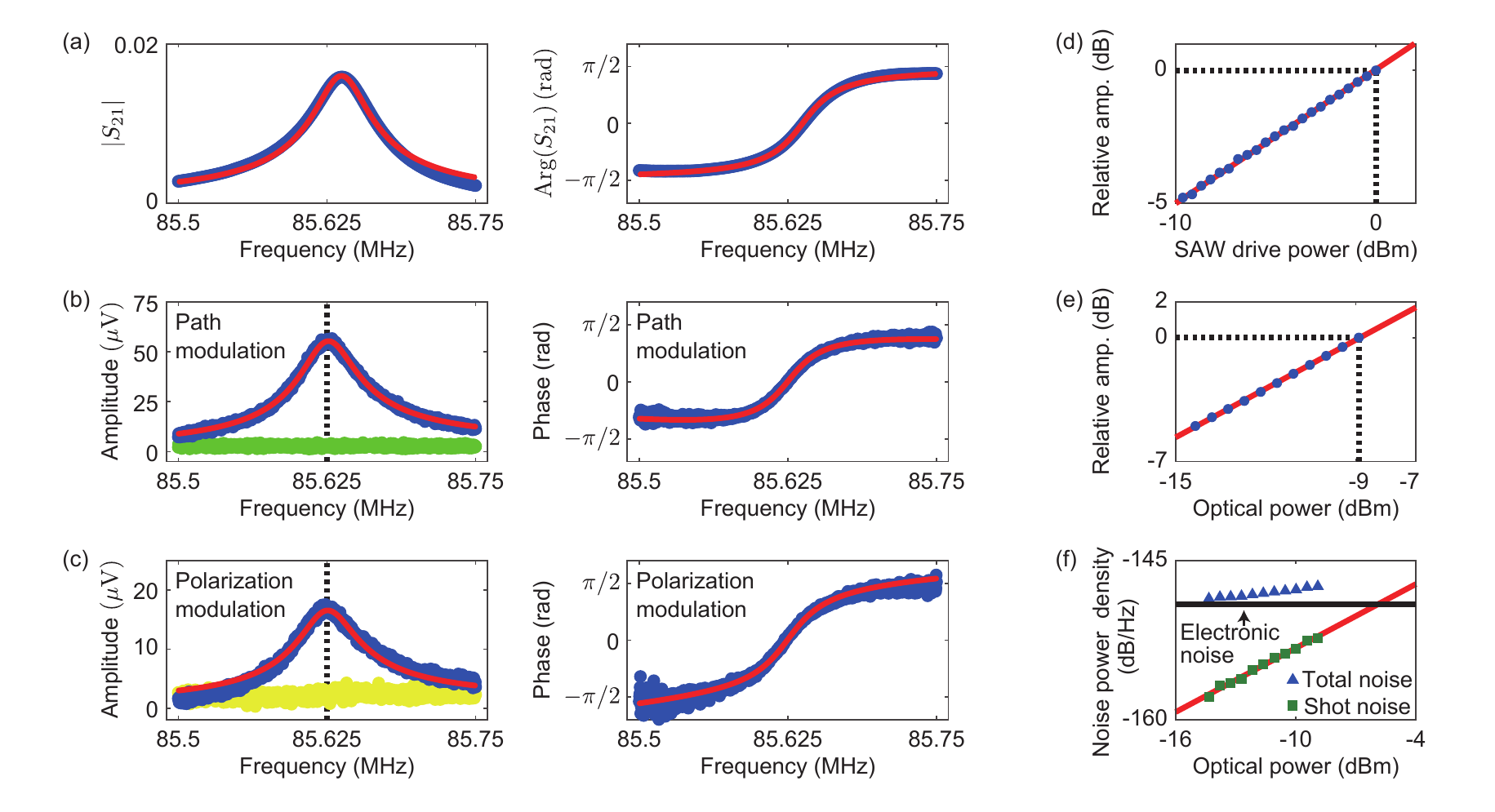}
\caption{
Spectra of the fundamental SAW resonator mode acquired by (a) electrical transmission measurement, (b) optical path modulation measurement, and (c) optical polarization modulation measurement. (a)~Rf transmission spectra (blue dots) from Port~1 to Port~2 with fitting curves (red line) . The peak around $85.63\,\mathrm{MHz}$ corresponds to the fundamental SAW resonator mode. The resonance frequency is almost the same as that in the following optical modulation spectroscopy. (b) Spectra of the optical path modulation signal (blue dots) and the background signal (green dots) with fitting curves (red line). (c) Spectra of the optical polarization modulation signal (blue dots) and the background signal (yellow dots) with fitting curves (red line). (d) Relative amplitudes for the optical path modulation (blue dots) as a function of the rf power that is used for driving the SAW with fitting curves (red line). (e) Relative amplitudes for the the optical path modulation (blue dots) as a function of the optical power $P_{i}$ into the balanced detector with fitting curves (red line). (f) Total noise power spectral density (blue triangles) and shot-noise power spectral density (green squares) at the measurement frequency $\omega/2\pi=85.625\,\mathrm{MHz}$ within the bandwidth $\Delta f =1\,\mathrm{kHz}$ as a function of the optical power $P_{i}$ into the balanced detector with fitting curves (red line). The shot-noise power is obtained by subtracting the separately measured electronic noise (black line) from the total noise. Note that we refer to the noise power as the square of the voltage measured by the lock-in amplifier, which is a quantity proportional to the actual noise power.
%221117
}
\label{fig:mainresults}
\end{center}
\end{figure*}

\section{Experiments} \label{sec:experiments}
%\subsection{experimental setup}
The experimental setup is schematically shown in Fig.~1(a).
A continuous-wave light with a wavelength of $1550\,\rm{nm}$ output from a laser (Santec: TSL-550) is polarized in the $z$-direction using a polarization beam splitter (PBS) and a half-wave plate (HWP), then focused by the objective lens (Mitutoyo: M-PLAN NIR 20X), and set so that the device surface is in focus.
The incident angle to the SAW device is set at $45^{\circ}$, as shown in Figs.~1(c) and 1(d).
We analyze the signal on the reflected light from the SAW device with two balanced detectors labeled (i) and (ii) in Fig.~1(a).
Those aim to detect (i) the optical path modulation and (ii) the optical polarization modulation~\cite{TH2021}, respectively.
The entire setup is operated at room temperature.

The optical system for (i)~the optical path modulation measurement is the part following the periscope-type mirror pair in Fig.~1(a).
The light split by the prism mirror is separately coupled to the photo-detectors of the balanced detector (Thorlabs: PDB415C).
This system works according to the measurement principle described in Sec.~\ref{sec:model} and is sensitive to the beam displacement $L$.
Note that several optical elements between the SAW device and the periscope-type mirror pair do not affect the optical path modulation.

In (ii)~the polarization modulation measurement system, the orthogonal polarization components of the reflected light are split by the PBS, coupled separately to multi-mode (MM) fibers, and differentially measured by the balanced detector (Thorlabs: PDB465C). Here, the polarization plane is rotated with the HWP after the SAW device by the angle of $45^{\circ}$ with respect to the original plane to maximize the amplitude of the polarization modulation~\cite{TH2021}. 
%221117

Figure~1(b) shows a microscope image of the SAW device.
The SAW device is a Fabry-Perot-type SAW resonator consisting of two Bragg mirrors, two interdigitated transducers (IDTs), and a single pad for optical observation formed by Ti($5$) and Au($80$) films on a $\LiNb$ substrate, where the values in parentheses are the thicknesses in nanometers.
Note that the single pad in the center of the device is in preparation for future experiments and is not significant in this paper.
The SAW resonator is fabricated so that the SAW propagation direction is along the crystalline $Z$-axis of the $\LiNb$ monocrystal.
The fundamental SAW resonator mode with a wavelength of $\sim\micron{40}$ is coherently generated by driving the IDT of Port~1 with an rf signal at the frequency of $\sim \MHz{86}$.
%221117

%\subsection{Results}
We first characterize the SAW resonator electrically.
Figure~2(a) shows the rf transmission spectra $|{S}_{21}|$ of the SAW resonator from Port~1 to Port~2 in Fig.~1(b) measured with a vector network analyzer.
By fitting with a model function, we find the resonance frequency of the fundamental SAW resonator mode is $\MHz{85.637}$ and internal quality factor is $2\times10^3$ (see Appendix~\ref{SM:S21} for details).
%221117

Figures 2(b) and 2(c) show the amplitude and phase spectra of (i) the optical path modulation and (ii) the optical polarization modulation.
These results are obtained when the optical spot (diameter is about $\micron{3}$) is placed at the same position in a rectangular Ti/Au pad region between the IDTs.
Both measurements are performed where the SAW resonator mode is driven from Port 1 with an rf signal with the power $P_{\mathrm{rf}}$ of $\dBm{-11}$ produced by mixing a fixed-frequency signal at $\MHz{80}$ from an rf generator (Rhode\&Schwarz: SGS100A) and a variable-frequency signal at $5.50$-$\MHz{5.75}$ from the lock-in amplifier (Zurich instruments: HF2LI).
The optical signals are acquired by the balanced detectors shown in Fig.~1(a), mixed-down with the $80$-$\rm{MHz}$ rf signal, and then demodulated with the lock-in amplifier.
The spectra in Figs.~2(b) and 2(c) show amplitude peaks and steep phase changes around $\MHz{85.6}$, which are similar to the rf transmission spectra in Fig.~2(a).
The red lines in Figs.~2(b) and 2(c) show the fitting results with a model function of the same type as the function used in Fig.~2(a) (see Appendix \ref{SM:S21} for details).
The agreement of these spectra is evidence that the rf signal is coherently converted into the optical path modulation signal and the polarization modulation signal via the SAW.
The agreement also indicates that the SAW can be observed simultaneously in both modulation measurements.
Note that the green dots in Fig.~2(b) are the background signal obtained when the optical beam is blocked in front of the balanced detector, and there is no noticeable structure.
We confirm that there is no stray rf field directly coupled to the balanced detector.
Note also that the yellow dots in Fig.~2(c) are the background signal obtained when the polarization plane of the reflected light from the SAW device are not rotated with respect to the original plane, and there is no noticeable structure.
This setting is insensitive to the optical polarization modulation.
Thus, we confirm that the observed signal (blue dots) is from the optical polarization modulation. 
%221117

Now, we measure the SAW drive power dependence and optical power dependence of the optical path modulation signal to verify the measurement principle described in Sec.~\ref{sec:model}.
Figure~2(d) shows the observed relative amplitude of the optical path modulation signal (blue dots) at the resonant frequency $85.625\,\mathrm{MHz}$ as a function of the SAW drive power from the lock-in amplifier. 
Here, the reference is the amplitude when SAW drive power is $0~\mathrm{dBm}$.
From the linear fitting (red line), we obtain the slope is $0.5$.
Substituting Eqs.~(\ref{eq:Tanduy}) and (\ref{eq:LandTheta}) into Eq.~(\ref{eq:VandL}), we find that the output voltage from the balanced detector $V(x,z,t)$ is proportional to the surface slope $\partial u_{y}(x,z,t)/\partial z$.  
And using the fact that the $y$-component of the surface displacement of a fundamental mode of the SAW resonator can be written as~\cite{TB2017},
\begin{equation}
u_{y}(x,z,t) = \alpha \cos(kz) \cos(\omega t), \label{eq:u_cavity}
\end{equation}
where $\alpha\cos(kz)$ is the amplitude, $k$ is the wave vector, and $\omega$ is the angular frequency, we can obtain $V(x,z,t)\propto \alpha$.
By considering that the amplitude of the displacement $|\alpha|$ is proportional to the square root of the SAW drive power, we can verify that the theoretical prediction and the experimental result agree.
Equation~(8) implies that the surface displacement also has the same dependence on the SAW drive power as in Fig.~2(d).
Furthermore, Fig.~2(e) shows the relative amplitude of the optical path modulation signal (blue dots) as a function of incident optical power going to the balanced detector. 
Here, the reference is the amplitude when optical power is $-9~\mathrm{dBm}$.
From the linear fitting, we obtain the slope equal to $1$.
By considering that the optical flux density $D$ in Eq.~(\ref{eq:VandL}) is proportional to the incident optical power, we can confirm that the theoretical prediction agrees with the experimental result.
%221117

Next, we evaluate the slope amplitude from the optical path modulation measurement result in Fig.~2(b) according to the scheme described in the last paragraph of Sec.~\ref{sec:model}.
Here, we describe a more specific procedure.
The first step is to obtain the optical shot noise level in the optical path modulation measurement system.
Figure~2(f) shows the total noise power spectral density (blue triangles) without SAW driving as a function of the optical power impinged into the balanced detector at the measurement rf frequency $\MHz{85.625}$ within the bandwidth $\Delta f =1\,\rm{kHz}$.
Since the optical shot noise and electronic noise are comparable in this measurement system, the shot noise (green squares) is obtained by subtracting the electronic noise (black line) from the total noise (blue triangles).
Note that the electronic noise is independent of the optical power as it mainly consists of thermal noise generated by the balance detector, cables, and the lock-in amplifier.
As shown by the red line in Fig.~2(f), the optical shot noise increases linearly with a slope of $1$ for the optical power, confirming the nature of the optical shot noise (see Appendix \ref{SM:PS} for details).
From the optical power detected when the optical spot is at a position $(x,z)$ on the SAW device, we can determine the shot-noise level corresponding to each optical position.
The next step is to calculate the ratio $R$ defined in Eq.~(\ref{eq:R}). By comparing the experimentally obtained signal power at $\omega/2\pi=\MHz{85.625}$ in Fig.~2(b) with the shot-noise level, we obtain the resultant ratio $R$ of $19.5\,\mathrm{dB}$.
Finally, substituting this value of ratio and the following parameters, $\omega_{0}=194\,\mathrm{THz}$, $\Delta f = 1\,\mathrm{kHz}$, $P_{i}= 0.13\,\mathrm{mW}$, $D=P_{i}/(6.1\,\mathrm{mm})$ (see Appendix \ref{SM:D} for details), and $F= 10\,\mathrm{mm}$, into Eq.~(\ref{eq:R}), we evaluate the slope amplitude $|{\theta}_n| =2.0\times 10^{-6}\,\mathrm{rad}$.
To validate this new method, we evaluate the slope amplitude in two different ways using literature values.
The first is an evaluation using the optical polarization modulation and the material complex refractive index, resulting in $|\theta_n| \sim 1.1\times10^{-6}\,\mathrm{rad}$ (see Appendix \ref{SM:Pol} for details).
The second is an evaluation using the SAW resonator mode volume and material constants, resulting in $|\theta_n| \sim 1.3\times10^{-6}\,\mathrm{rad}$ (see Appendix \ref{SM:volume} for details).
All of the evaluation values are very close to each other, confirming the validity of the optical path modulation measurement.
%221118

%%%% Figure 3%%%%
\begin{figure}[!]
\begin{center}
\includegraphics[width=8.6cm,angle=0]{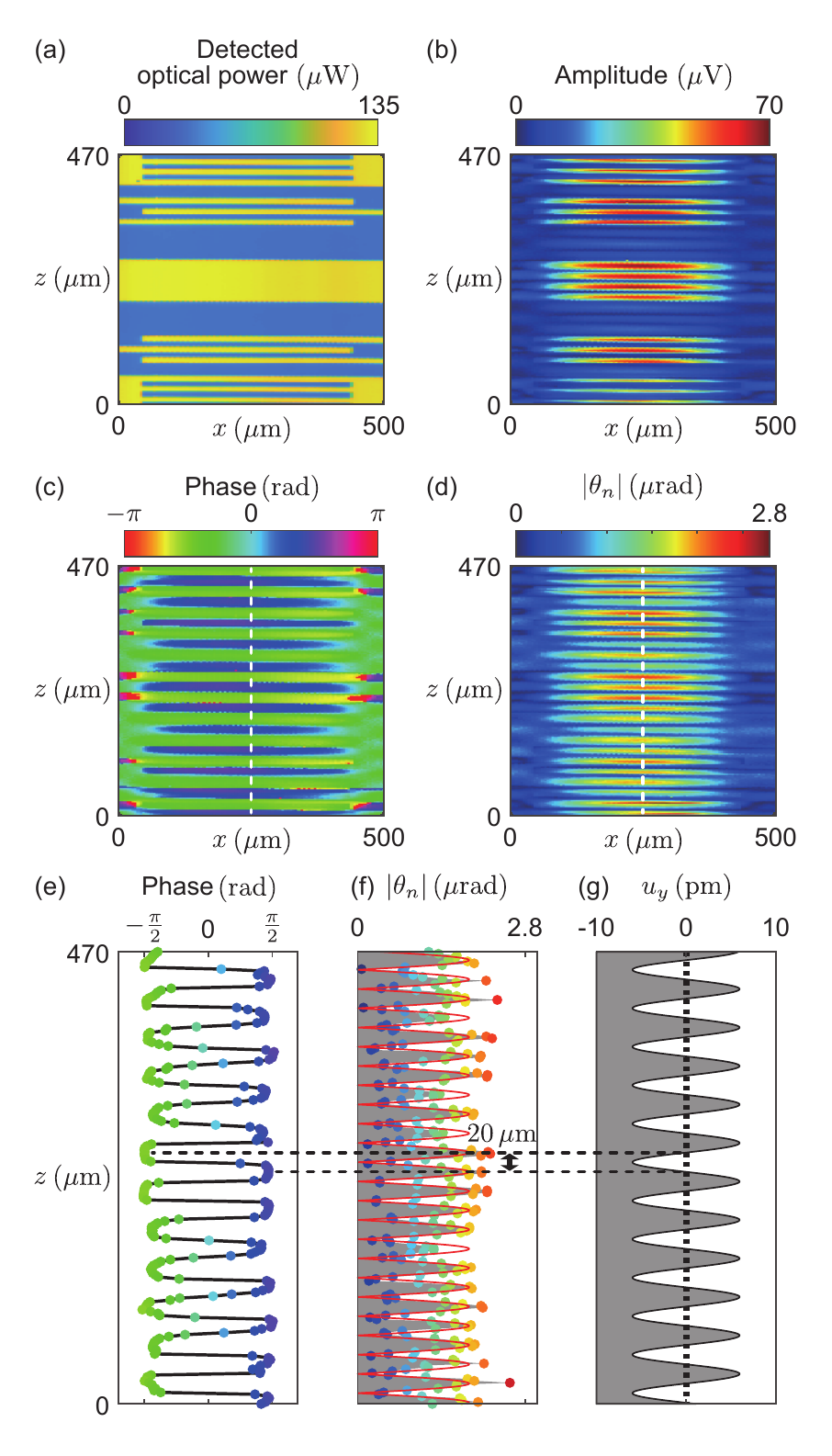}
\caption{
Imaging results of the SAW resonator by optical path modulation measurement. Color maps of (a) detected optical power into the balanced detector, (b) amplitude, and (c) phase acquired by the lock-in amplifier at a frequency $\omega/2\pi=85.625\,\mathrm{MHz}$, and (d) evaluated slope amplitude as a function of the position of optical spot. Together with the color maps, cross-sections of (e) the phase and (f) the slope amplitude at the position $x=250\,\mathrm{\mu m}$ (indicated by white dash lines in the color maps) are shown.   
The color of dots in (e) and (f) is consistent with the color in (c) and(d), respectively.
(g) The $y$-component of the displacement at maximum displacement estimated using the fitting result (red line) in~(f).
}
\label{fig:mainresults2}
\end{center}
\end{figure}

To gain further insight, we demonstrate the acquisition of two-dimensional plots of the amplitude and the phase of the surface slope in a region ($470\,\rm{\mu m}\times 500\,\rm{\mu m}$) around center of the SAW device .
Figures~3(a), 3(b), and 3(c) show the two-dimensional plots of the detected optical power and the amplitude and the phase of the optical path modulation signal while the optical spot position is scanned in the $x$ and $z$ axes every $\micron{1}$.
Figure~3(a) reflects the reflectance at each optical position, thus confirming the surface structure of the SAW device in the scanned area.
Figure 3(d) shows a color map of the slope amplitude~$|{\theta}_{n}(x,z)|$ at each optical position evaluated according to the specific procedure described in the previous paragraph. 
Figure~3(e) [3(f)] is a cross section of Fig.~3(c) [3(d)] at the position $x=250\,\mathrm{\mu m}$.
The interval between the peak positions of $|{\theta}_{n}(z)|$ in Fig.~3(f) is about $20\,\mathrm{\mu m}$, half the wavelength of the excited SAW, and the phase in Fig.~3(e) changes by exactly $\pi$ at each peak.
Furthermore, in Figs.~3(c) and 3(d), there are no nodes in the $x$-direction for both phase and amplitude.
These results show that the observed SAW mode is the fundamental mode of the SAW resonator.

Finally, we estimate the $y$-component of the surface displacement $u_{y}(x,z)$ from the obtained slope amplitude.
The surface slope of the fundamental mode of the SAW resonator can be written explicitly as
\begin{equation}
    \theta_{n}(x,z,t) = -\alpha k \sin(kz) \cos(\omega t),
\end{equation}
which is obtained by substituting Eq.~(8) into Eq.~(1).
From fitting the surface slope in Fig.~3(f) using Eq.~(9) (red line), we obtain $\alpha=6.0\,\mathrm{pm}$ and $k=0.16\,\mathrm{rad/\mu m}$, showing that the maximum displacement is $6.0\,\mathrm{pm}$.
By substituting these results into Eq.~(8), we can estimate the surface displacement at maximum displacement, as in Fig.~3(g).

\section{Discussion} \label{sec.discussion}
The optical path and polarization modulation measurements are helpful in quantitatively observing the surface slope due to SAW.
When a measurement system is in the shot-noise-limited regime, both measurements can be calibrated with optical shot noise, and the amplitude of the surface slope can be evaluated.
In practice, the optical path modulation measurement is more suitable for the surface slope measurement since it does not require information on the material complex refractive index at an optical spot~\cite{TH2021}, and the alignment of the optical system is relatively simple.

Furthermore, the surface slope can be used to estimate the surface displacement, the most natural physical quantity that characterizes SAW.
Although this paper demonstrates the method's validity using the SAW in the Fabry-Perot-type resonator, the method also applies to propagating SAW, which is used in many studies and applications.
In the case of propagating SAW, the surface slope can be written as in Eq.~(5).
The equation shows that the slope amplitude ($|\theta_n(x,z)|=|\tilde\theta_n(x,z)|$) is constant within negligible propagating loss, and the phase evolves linearly in the propagation direction.
Moreover, the $y$-component of the displacement of the propagating SAW can be written as $u_y(x,z,t)=\beta\exp\lbrack i(kz-\omega t)\rbrack$, and by substituting into Eq.~(1), we obtain $|\tilde\theta_n(x,z)|=|\beta k|$.
The displacement amplitude $\beta$ can be estimated by substituting the experimentally evaluated slope amplitude $|\tilde\theta_n(x,z)|$ and $k$ into the above equation.
Thus, the method enables the acquisition of spatial information of the displacement $u_y(x,z,t)$ for both resonator SAW and propagating SAW.
Its acquisition will contribute to further developing phenomena with SAW~\cite{W2002,WD2011,WH2012,GM2015,BV2020,MC2012,PB2015,PS2015,M2013,IMM2014,K2017,K2020,SK2021,B2022}, as it allows quantitative estimation of, for example, strain, electric field, vorticity, and the angular momentum distribution associated with SAW.

Here, we discuss the influence of the optoelastic effect on this optical path modulation measurement. 
In this paper, the surface slope is derived from the optical path modulation measurement and the theoretical method using the mode volume. 
Neither method incorporates the optoelastic effect. The fact that the surface slopes evaluated by those two methods are almost identical indicates that the influence of the optoelastic effect is negligible in this experimental situation. 
Furthermore, the result in Fig.~3(f) shows that the magnitudes of the calculated surface slopes do not differ between the Au and $\mathrm{LiNbO_3}$ regions, which have different optoelastic constants. 
The fact is another evidence that the optoelastic effect does not contribute to this experiment.
Note that the optical path modulation measurement cannot be applied to SAW without surface waving, such as a shear horizontal wave. The optoelastic effect may be the key to imaging such types of SAW.

Let us now discuss the limitations of the optical path modulation measurement.
First, let us discuss spatial resolution.
The optical spot size determines the spatial resolution of this method.
The period of the slope should be sufficiently larger than the spot size, but in practice, it may be applicable up to the same size.
If a visible laser with a wavelength of $532\,\mathrm{nm}$ is used instead of the laser with a wavelength of $1.5\,\mathrm{\mu m}$ used in this paper, the spot size can be reduced to less than $1\,\mathrm{\mu m}$ in diameter.
On the $\mathrm{LiNbO_3}$ substrate, which is a typical substrate for SAW device, this method can be applied to SAW up to about $1.7\,\mathrm{GHz}$ for propagating SAW and $0.8\,\mathrm{GHz}$ for SAW in Fabry-Perot-type resonator.

Next, for the case where SAWs with two or more wave vectors are superposition.
In this paper, we focused only on the surface slope in the $z$-direction shown in Fig.~1(d), but the slope in the $x$-direction can also be observed by appropriately changing the optical beam displacement measurement basis.
Thus, the two-dimensional surface slope due to SAW can be observed by combining the optical path modulation measurements with orthogonal measurement bases.

Finally, although we limited our discussion to SAW as the measurement target, the shot-noise-based calibration method proposed in this paper can also be used to easily quantify optical displacement measurements of cantilevers and torsion pendulums, which were used in many studies~\cite{WH2021,RB2004}.
%221118

\section{Conclusion}
In summary, we demonstrated the optical path modulation measurement of SAW.
The precisely calibrated optical path modulation measurement can be a practical tool to quantitatively evaluate the spatiotemporal profile of the displacement field of SAW.
It will help quantitatively calculate the various physical quantities associated with SAW.
By assigning the optical polarization modulation measurement to magnon detection via the magneto-optic Kerr effect, simultaneous quantitative measurements of the displacement profile of SAW and the magnetization profile of the magnon may become feasible.
Those results will lead to an experimental determination of the coupling strength of the phonon-magnon coupling.

\acknowledgments
We would like to thank Koji~Usami, Rekishu~Yamazaki, Yuichi~Ohnuma, Alto~Osada, Atsushi~Noguchi, and Toshiyuki~Hosoya for useful discussions. 
We acknowledge financial support from JST PRESTO (Grant No. JPMJPR200A) and JSPS KAKENHI (Grant No. JP22K14589).
R.~S. and K.~T. are supported, in part, by the Japan Society for the Promotion of Science.

\appendix
\section{Theoretical expression of power spectrum for the optical path modulation}\label{SM:PS}
In sec.~\ref{sec:model}, we describe the mechanism of the optical path modulation as a result of dynamic surface slope variation due to the SAW. 
This appendix describes how to obtain the power spectrum represented by Eq.~(\ref{eq:PS}).

\begin{figure}[t]
\begin{center}
\includegraphics[width=8.6cm,angle=0]{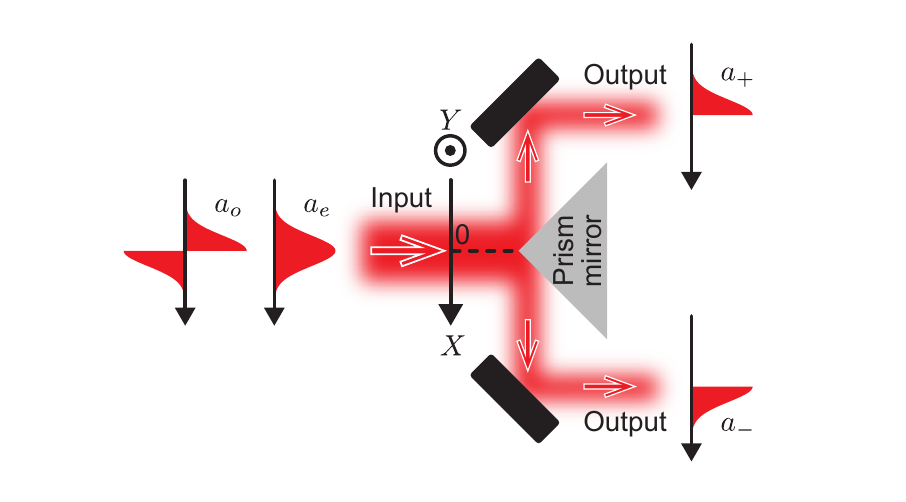}
\caption{
Schematic illustration of spatial modes of light split by the prism mirror.
}
\label{fig:app1}
\end{center}
\end{figure}

To begin with, we assume for simplicity that a Gaussian beam is impinging on the center of the prism mirror in Fig.~1(a).
Figure~\ref{fig:app1} shows spatial modes of the light entering the prism mirror and the light reflected by the prism mirror.
Now, we introduce an odd-Gaussian mode that is antisymmetric to the $Y$-axis~\cite{BF2003}. The relationship of the electric fields of the input and reflected light can be written as
\begin{equation}
\left[
\begin{array}{c}
a_{+}(t)\\
a_{-}(t)
\end{array}
\right]
=
\underbrace{\frac{1}{\sqrt{2}}
\left[
\begin{array}{cc}
1  & 1  \\
1 & -1 \\
\end{array}
\right]}_{B}
\left[
\begin{array}{c}
a_{e}(t) \\
a_{o} (t)\\
\end{array}
\right], \label{eq:u_matrix}
\end{equation} 
where $a_{e}(t)$, $a_{o}(t)$, and $a_{\pm}(t)$ are electric fields of the Gausissian beam, the odd-Gaussian beam, and the beam reflected in the $\pm X$-direction, respectively.
Since there is no change in optical power between the input side ($a_{e}^{*}a_{e}+a_{o}^{*}a_{o}$) and output side ($a_{+}^{*}a_{+}+a_{-}^{*}a_{-}$), matrix $B$ must be unitary.
Any two-dimensional unitary matrix can be expressed in the following form
\begin{equation}
\left[
\begin{array}{cc}
\cos{\phi}  & \sin{\phi}  \\
\sin{\phi} & -\cos{\phi} \\
\end{array}
\right], \label{eq:BS_matrix}
\end{equation} 
where $\phi$ is an arbitrary real number.
Hence the matrix $B$ is the case where $\phi = \pi/4$.
%221129

When the SAW is driven, the optical displacement in the $X$-direction occurs, which subsequently modulates the optical power split by the prism mirror.
In practice, the amount of the optical displacement $L$ is much smaller than the optical beam width, thus, we can ignore the change in spatial mode and consider that only the matrix $B$ is modulated.
Under this assumption, the input-output relation with SAW driven can be written as
\begin{equation}
\left[
\begin{array}{c}
a_{+}(t)\\
a_{-}(t)
\end{array}
\right]
=
\underbrace{\frac{1}{\sqrt{2}}
\left[
\begin{array}{cc}
(1-\Delta \phi(t))  & (1+\Delta \phi(t))  \\
(1+\Delta \phi(t)) & -(1 -\Delta \phi(t)) \\
\end{array}
\right]}_{B'}
\left[
\begin{array}{c}
a_{e}(t) \\
a_{o}(t) \\
\end{array}
\right]. \label{eq:d_matrix}
\end{equation} 
The matrix $B'$ is obtained by Taylor expanding the contents of Eq.~(\ref{eq:BS_matrix}) around $\phi=\pi/4$ and neglecting higher-order terms of the tiny quantity $\Delta \phi(t)$. $\Delta \phi(t)$ is explicitly written as
\begin{equation}
    \Delta \phi(t) =\frac{L(t)\times D}{P_{i}}. \label{eq:def_phi}
\end{equation}
In Sec.~\ref{sec:model}, we include $x$ and $z$, which indicate the optical spot position, in the arguments of physical quantities such as $L$. In this appendix, they are not explicitly indicated as arguments hereafter either, since they are considered fixed.
%221129

%%%% Figure Appendix 1%%%%

In the (i) optical path modulation measurement system in Fig.~1(a), we measure the instantaneous output voltage from the balanced detector.
The output voltage is proportional to the difference between the optical power observed by the two photodetectors,
\begin{equation}
V(t)\propto \braket{a_{+}^{*}(t) a_{+}(t)-a_{-}^{*}(t) a_{-}(t)}.
 \label{eq:def_V_classic}
\end{equation}
For the shot-noise-based calibration scheme, we need to deal with the expected value of the optical power difference and the fluctuations that include the contribution of the shot noise (vacuum noise).
To this end, we convert the electric field in Eq.~(\ref{eq:d_matrix}) into quantum mechanical annihilation operators by the following simple prescription:
\begin{equation}
\left[
\begin{array}{c}
a_{e}(t)\\
a_{o}(t)\\
a_{+}(t)\\
a_{-}(t)\\
\end{array}
\right]
\longrightarrow
\left[
\begin{array}{c}
\hat{a}_{e}(t)\\
\hat{a}_{o}(t)\\
\hat{a}_{+}(t)\\
\hat{a}_{-}(t)\\
\end{array}
\right].
 \label{eq:operators}
\end{equation} 
We assume that these time-domain operators of light satisfy the commutation relation, $[\hat{a}_{j}(t), \hat{a}^{\dagger}_{j^{\prime}}(t^{\prime})]=\delta_{jj^{\prime}}\delta(t-t^{\prime})$.
The output voltage $\hat{V}(t)$ can then be expressed as
\begin{eqnarray}
\hat{V}(t)\propto&& \left(\hat{a}_{\mathrm{+}}^{\dagger}(t) \hat{a}_{\mathrm{+}}(t) - \hat{a}_{\mathrm{-}}^{\dagger}(t) \hat{a}_{\mathrm{-}}(t)\right) \notag\\
=&& \frac{1}{\sqrt{2}}\Biggl[\Biggl(\lbrack1-\Delta \phi(t)\rbrack\hat{a}_{e}^{\dagger}(t) +
\lbrack1+\Delta \phi(t)\rbrack\hat{a}_{o}^{\dagger}(t)\Biggr) \notag\\
&&\Biggl(\lbrack1-\Delta \phi(t)\rbrack\hat{a}_{e}(t) +
\lbrack1+\Delta \phi(t)\rbrack\hat{a}_{o}(t)\Biggr) \notag\\
&&-
\Biggl(\lbrack1+\Delta \phi(t)\rbrack\hat{a}_{e}^{\dagger}(t) -
\lbrack1-\Delta \phi(t)\rbrack\hat{a}_{o}^{\dagger}(t)\Biggr) \notag\\
&&\Biggl(\lbrack1+\Delta \phi(t)\rbrack\hat{a}_{e}(t) -
\lbrack1-\Delta \phi(t)\rbrack\hat{a}_{o}(t)\Biggr)\Biggr].
\label{eq:def_V_quantum}
\end{eqnarray}
%221129

Assuming further that the input light is a coherent state of the annihilation operator $\hat{a}_{e}(t)$, the operator $\hat{a}_{e}$ can be split into a classical part $\beta_{e}$ and a part representing the quantum fluctuation $\hat{d}_{e}(t)$ as
\begin{equation}
\hat{a}_{e}(t) = \beta_{e} + \hat{d}_{e}(t).
\end{equation}
Here, $\beta_{e}$ is related to the power $P_i$ of the input light as $|\beta_{e}|^2 = \frac{P_i}{\hbar \omega_0}$.
Hereafter we will assume that the classical part $\beta_{e}$ is real for simplicity.
Note that another annihilation operator $\hat{a}_{o}$ has only part representing quantum fluctuations.
Based on Eq.~(\ref{eq:def_V_quantum}), the auto-correlation of $\hat{V}(t)$ is given by
\begin{eqnarray}
\braket{\hat{V}(t) \hat{V}(t+\tau)} \propto && 
4\beta_{e}^{4} \underbrace{\langle \Delta\phi(t) \Delta\phi(t+\tau) \rangle}_{2D^2 F^2 |{\theta}_{n}|^{2}\cos (\omega_{\mathrm{SAW}}\tau)/P_{i}^{2}} \notag\\
&&+ \beta_{e}^{2}\underbrace{\langle\hat{a}_{o}(t) \hat{a}_{o}^{\dagger}(t+\tau)\rangle}_{\delta(\tau)}, \label{eq:ac_V}
\end{eqnarray}
where the auto-correlations of the operators for the light are evaluated concerning the vacuum states.
As for the auto-correlation of $\Delta\phi(t)$ in Eq.~(\ref{eq:ac_V}) we take the form given by Eqs.~(\ref{eq:LandTheta}), (\ref{eq:theta_xzt}), and (\ref{eq:def_phi}) and evaluate its auto-correlation. Consequently, we have
\begin{eqnarray}
\braket{\hat{V}(t) \hat{V}(t+\tau)} \propto && 
\frac{8\beta_{e}^4D^2F^2|{\theta}_{n}|^2}{P_{i}^2}\left(e^{i\omega_{\mathrm{SAW}}\tau}+e^{-i\omega_{\mathrm{SAW}}\tau}\right) \notag\\
&&\,\,\,\,\,\,\,\,\,\,\,\,\,\,\,\,\,\,+ \beta_{e}^{2}\delta(\tau). \label{eq:ac_V_rev}
\end{eqnarray}
Fourier transforming $\braket{\hat{V}(t) \hat{V}(t+\tau)}$, we obtain the following power spectral density $S_{VV}(\omega)$:
\begin{eqnarray}
    S_{VV}(\omega)\propto&&\frac{8\beta_{e}^4D^2F^2|{\theta}_{n}|^2}{P_{i}^2}\lbrack 2\pi\delta(\omega-\omega_{\mathrm{SAW}})+2\pi\delta(\omega+\omega_{\mathrm{SAW}})\rbrack\notag\\
&&+ \beta_{e}^{2}. \label{eq:SVV}
\end{eqnarray}
At resonance $\omega=\omega_{\mathrm{SAW}}$, the power within the bandwidth $\Delta f \equiv \frac{\Delta \omega}{2 \pi}$ reads,
\begin{eqnarray}
S_{VV}(\omega_{\mathrm{SAW}}) \Delta f \propto \underbrace{\frac{8 D^2 F^2 |{\theta}_n|^2}{\hbar^2 \omega_0^2}}_{\mathrm{signal}} + \underbrace{\frac{P_{i}}{\hbar \omega_{0}} \Delta f}_{\mathrm{shot\,noise}},    \label{eq:Svv_comp}
\end{eqnarray}
where the first term is the signal due to the coherent SAW excitation and the second term is the frequency-independent shot noise.
%%221129

\section{Evaluation of $D$}\label{SM:D}
\begin{figure}[t]
\begin{center}
\includegraphics[width=8.6cm,angle=0]{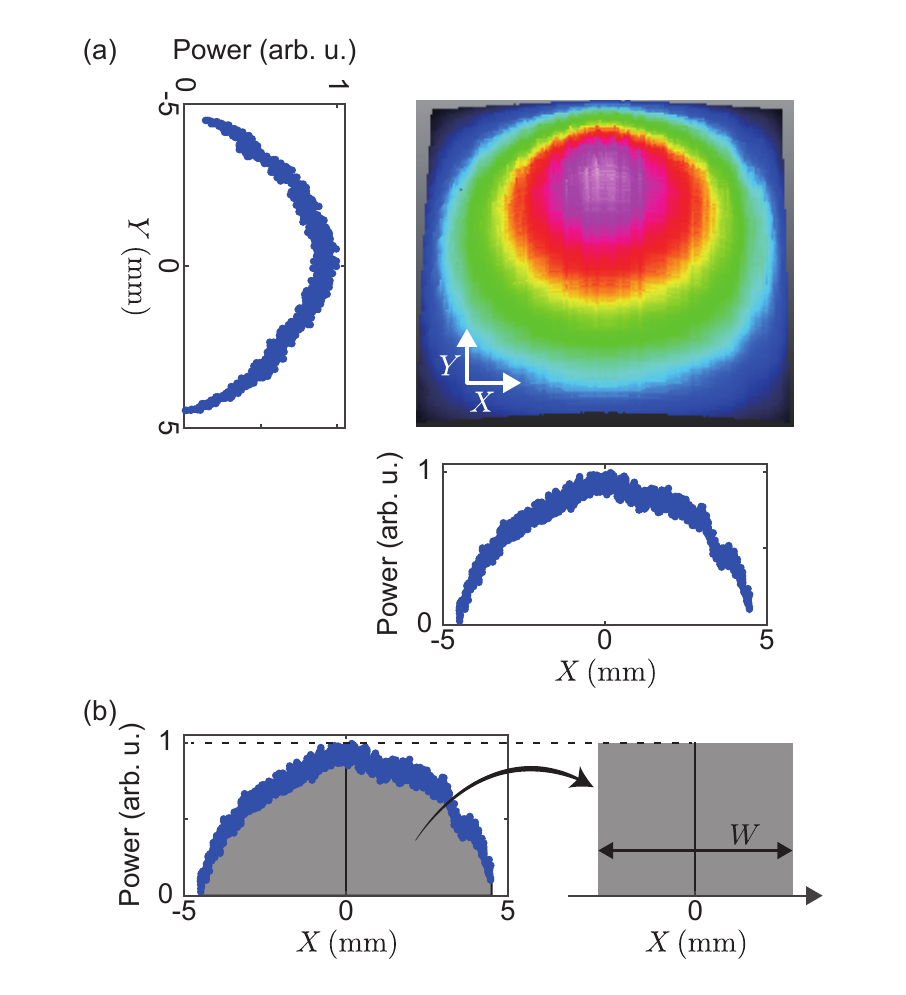}
\caption{
(a) Three-dimensional and cross-sectional optical power profiles. (b) Profile in the $X$-direction [same as lower panel in (a)] and the profile transformed into a rectangle, with the area and top height constant. }
\label{fig:app2}
\end{center}
\end{figure}

The optical power density $D$ in the $X$-direction around the center of the incident light to the prism mirror in Fig.~1(a) can be independently evaluated.
We place a beam profiler (Thorlabs: BP209-IR/M) in front of the prism mirror.
Figure~\ref{fig:app2}(a) shows a measured incident light profile.
The result indicates that the light is not a Gaussian beam.
Since we confirmed that the light transmitted through and focused by the first objective lens is a Gaussian beam, this result may be due to truncation by the pupil of the second objective lens.
Note that the theory described in Appendix~\ref{SM:PS} is valid even if the light input to the prism mirror is not a Gaussian beam.

%%%% Figure Appendix 2%%%%

To obtain the optical power density $D$ around the center of the beam, we transform the cross-section in the $X$-direction in the lower panel in Fig.~\ref{fig:app2}(a) into a rectangle with area and top height constant, as shown in Fig.~\ref{fig:app2}(b).
From the width $W=6.1\,\mathrm{mm}$ of the rectangle and the actual optical power $P_{i}$, the optical power density $D$ is given by $D=P_{i}/W=P_{i}/\left(6.1\,\mathrm{mm}\right)$.
%221129

\section{Transmission coefficient $\mathrm{S_{21}}$ \\of the SAW resonator mode }\label{SM:S21}
\begin{figure}[t]
\begin{center}
\includegraphics[width=8.6cm,angle=0]{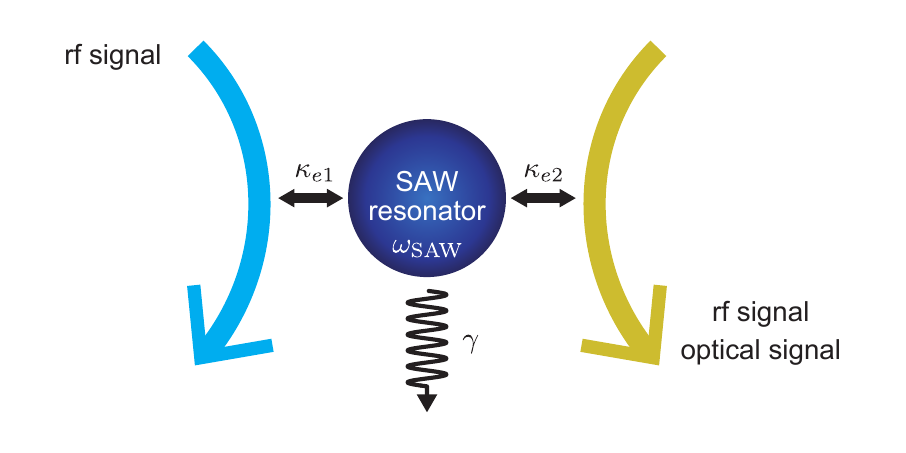}
\caption{
Physical model of the system in which signals are transformed via SAW resonator mode.
}
\label{fig:coupledsystem}
\end{center}
\end{figure}

The rf signal input from Port 1 in Fig.~1(b) is converted via the SAW resonator mode into an rf signal output to Port~2 and optical signals.
All of those phenomena can be described by a physical model in Fig.~\ref{fig:coupledsystem}~\cite{CD2010,HO2016}.
Here, $\omega_{\mathrm{SAW}}$ is a resonant angular frequency of the SAW resonator mode, $\kappa_{e1}$ ($\kappa_{e2}$) is the coupling rate between the input (output) field and the SAW resonator mode, and $\gamma$ is the internal loss rate of the SAW resonator mode.

In the case of conversion between rf signals, $\kappa_{e1}$ and $\kappa_{e2}$ represent the coupling rates due to the piezoelectric effect between the SAW and rf signal.
The transmission coefficient ${S}_{21}$ obtained by the network analyzer can be written as
\begin{equation}
    {S}_{21}(\omega)= \frac{\sqrt{\kappa_{e1} \kappa_{e2}}}{i(\omega-\omega_{\mathrm{SAW}})+\frac{\kappa_{e1}+\kappa_{e2}+\gamma}{2}}.\label{eq:S21}
\end{equation}
From the fitting using Eq.~(\ref{eq:S21}) in Fig.~2(a) we deduce $\kappa_{e1}/2\pi=\kappa_{e2}/2\pi=370\,\mathrm{Hz}$, $\omega_{\mathrm{SAW}}/2\pi=85.64\,\mathrm{MHz}$, and $\gamma/2\pi=46\,\mathrm{kHz}$.
Here, we assume that $\kappa_{e1}$ and $\kappa_{e2}$ are equal because the two IDTs have the same geometry and position in the SAW resonator.
From this result, the internal quality factor of the SAW resonator mode is determined to be $2\times10^3$.

%%%% Figure Appendix%%%%

In converting the rf signal to optical signals, $\kappa_{e1}$ still represents the coupling rate between the SAW resonator mode and rf signal due to the piezoelectric effect.
On the other hand, $\kappa_{e2}$ represents the coupling rate between the SAW resonator mode and optical signals due to the boundary effects during reflection at the SAW device surface.
The voltage spectrum $K(\omega)$ when the rf signal is converted to the optical path modulation signal or the polarization modulation signal can be written as
\begin{equation}
    K(\omega)= \frac{AB}{i(\omega-\omega_{\mathrm{SAW}})+B}. \label{eq:T}
\end{equation}
Here, we do not require to specify the coupling and loss rates. Thus we simplify Eq.~(\ref{eq:S21}).
From the fitting using Eq.~(\ref{eq:T}) in Fig.~2(b) [2(c)], we deduce $A=52.5 \,\mathrm{\mu V}\,(17.7\,\mathrm{\mu V})$, $B/2\pi=24\,\mathrm{kHz}\,(25\,\mathrm{kHz})$, and $\omega_{\mathrm{SAW}}/2\pi=85.626 \,\mathrm{MHz}\,(85.625\,\mathrm{MHz})$.
The fact that both amplitude and phase in Figs.~2(b) and 2(c) are fitted well is evidence that the rf signal is coherently converted to optical signals via the SAW resonator mode.
%%221129

\section{Evaluation of $|{\theta}_{n}|$ with the optical polarization modulation}\label{SM:Pol}
In this appendix, we evaluate the slope amplitude $|{\theta}_{n}|$ from the optical polarization modulation measurement in Fig.~2(c),
following the previously reported method~\cite{TH2021}.
Defining $R_{\mathrm{pol}}$ as the ratio of the power of the optical polarization modulation signal to the power of the optical shot noise, $R_{\mathrm{pol}}$ can be written as
\begin{equation}
R_\mathrm{pol} = \frac{2 (r_{s} - r_{p})^2 P_{i}^{\mathrm{pol}} |\theta_n|^2}{\hbar \omega_{0} \Delta f}. \label{eq:pol_SNR}
\end{equation}
Here, $P_{i}^{\mathrm{pol}}$ is the sum of the optical power entering the two photodetectors constituting the balanced detector for polarization modulation detection.
$r_{s}$ and $r_{p}$ are the Fresnel reflection coefficients for the $s$- and $p$-polarized light at an incident angle of $\pi/4$.
The quantities other than $|{\theta}_{n}|$ are experimentally independently evaluable values or literature values.

%%%% Figure App3%%%%
\begin{figure}[b]
\begin{center}
\includegraphics[width=8.6cm,angle=0]{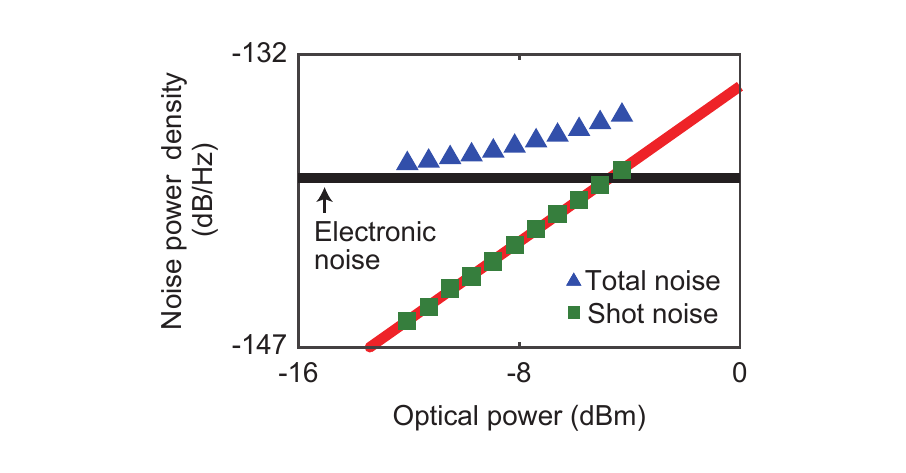}
\caption{
Total noise power density (blue triangles) and shot-noise power density (green squares) as a function of the optical power $P_{i}^{\mathrm{pol}}$ at a measurement frequency $\omega/2\pi=85.625\,\mathrm{MHz}$. The shot-noise power density is obtained by subtracting separately measured electronic noise power density (black line) from the total noise power density. Note that we refer to the noise power as the square of the voltage measured by the lock-in amplifier, which is a quantity proportional to the actual noise power.
}
\label{fig:app_pol_shotnoise}
\end{center}
\end{figure}

\begin{table}[t]
  \begin{center}
    \caption{Optical parameters of Au~\cite{JC1972}.} 
    \begin{tabular}{|c||c|} \hline
      Parameter & Au   \\ \hline \hline
      Refractive index & $0.524+ 10.7 \,i$  \\ \hline 
      $r_p$ with $\theta_1=\pi/4$ & $0.953+0.257 \,i$ \\ \hline
      $r_s$ with $\theta_1=\pi/4$& $-(0.985+0.130 \,i)$ \\ \hline
    \end{tabular}
  \label{table:Optical_parameters}
  \end{center}
\end{table}

Figure~\ref{fig:app_pol_shotnoise} shows the total noise level (blue triangles) of the polarization modulation measurement system as a function of the optical power $P_{i}^{\mathrm{pol}}$ at the measurement frequency $\omega/2\pi=85.625\,\mathrm{MHz}$ within the bandwidth $\Delta f =1\,\mathrm{kHz}$.
The shot-noise level and the electronic-noise level are comparable.
The shot-noise level (green squares) is obtained by subtracting the electronic noise (black line) from the total noise (blue triangles).
The shot-noise level grows linearly with the optical power $P_{i}^{\mathrm{pol}}$ as indicated by the red line (the slope of the line is $1.0$).
This result is in perfect agreement with the theoretical expression in Eq.~(\ref{eq:Svv_comp}).
From the observed optical power $P_{i}^{\mathrm{pol}}=0.31\,\mathrm{mW}$, we can assign a shot-noise level in the polarization modulation measurement system.

From the results in Figs.~2(c) and \ref{fig:app_pol_shotnoise}, the resultant $R_{\mathrm{pol}}$ is~$13.6\,\mathrm{dB}$.
Using Eq.~(D1), the material-dependent parameters listed in Table \ref{table:Optical_parameters}, and the following parameters $\omega_{0}/2\pi=194\,\mathrm{THz}$ and ${\Delta f}=1\,\mathrm{kHz}$, we then obtain the slope amplitude
\begin{equation}
    |{\theta}_{n}|\sim1.1\times10^{-6}\,\mathrm{rad}.
\end{equation}
%%221129

\section{Evaluation of $|{\theta}_{n}|$ with the SAW resonator mode volume}\label{SM:volume}
In this appendix, we evaluate the slope amplitude $|{\theta}_{n}|$ from the effective mode volume of the SAW resonator mode.
For simplicity, we only consider the surface slope incurred by the displacement of the $\mathrm{LiNbO_3}$ substrate.
Since the thickness of the Ti/Au film ($85\,\mathrm{nm}$) is far thinner compared with the SAW wavelength ($40\,\mathrm{\mu m}$), considering only the displacement of the bare $\mathrm{LiNbO_3}$ would be warranted.

The amplitude of the displacement field $\tilde{u}_y(x,z)$ of the SAW resonator mode can be obtained by $\tilde{u}_y(x,z)$ as a displacement of a simple harmonic oscillator with the zero-point fluctuation $U_0$~\cite{SK2015}, that is,
\begin{equation}
    \tilde{u}_y(x,z)\sim U_0 \sqrt{N} \sin(kz).\label{eq:uy}
\end{equation}
Here, $N$ is the number of phonons excited in the effective mode volume $V$ and $k=2\pi/\lambda_{\mathrm{SAW}}$ is the wave vector.
The effective mode volume is defined as $V=w\times l \times d$, where $w$ is the width of the IDT, $l$ is the effective length in the SAW propagation direction, and $d$ is the effective depth of the SAW resonator mode. Since the SAW is localized on the order of SAW wavelength in the depth direction~\cite{TB2017}, we assume $d\sim \lambda_{\mathrm{SAW}}$.

%%%% Figure App%%%%
\begin{figure}[t]
\begin{center}
\includegraphics[width=8.6cm,angle=0]{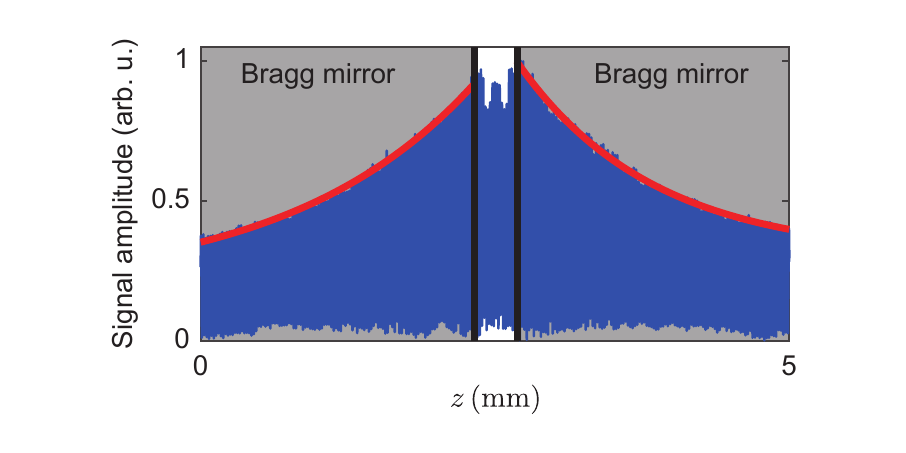}
\caption{
Position dependence of the optical path modulation signal induced by the SAW.
The blue line represents the signal amplitude measured at a measurement frequency of $85.625\,\mathrm{MHz}$ and normalized by the reflectance at each optical position.
The red line represents the results of exponential fittings of its envelope in the Bragg mirror regions (enclosed by the gray rectangle).
The penetration lengths defined by the length at which the signal decreases by $1/e$ are about $1.4\,\mathrm{mm}$ for each, where $e$ is Euler's number.
Note that here the SAW is excited through the IDT of Port~1.
}
\label{fig:imaging}
\end{center}
\end{figure}

\begin{table}[b]
  \begin{center}
    \caption{Parameters of SAW resonator.} 
    \begin{tabular}{|c||c|} \hline
      Parameter & Value   \\ \hline \hline
      Frequency: $\omega_{\mathrm{SAW}}/2\pi$ & $85.625\,\mathrm{MHz}$  \\ \hline 
      IDT width: $w$  & $0.4\,\mathrm{mm}$\\ \hline
      Effective length: $l$  & $3.1\,\mathrm{mm}$ \\ \hline
      SAW wavelength: $\lambda_{\mathrm{SAW}}$ &  $40\,\mathrm{\mu m}$ \\ \hline
    \end{tabular}
  \label{table:SAW_parameters}
  \end{center}
\end{table}

To experimentally determine the effective length $l$, we perform imaging of the SAW resonator over a long range using the optical path modulation.
Figure~\ref{fig:imaging} shows the position dependence of the optical path modulation signal normalized by the reflectance for each optical position at $x=250\,\mathrm{\mu m}$ in Fig.~\ref{fig:mainresults2}(a).
The black lines represent the positions of the Bragg mirror edges, and the red lines represent exponential fittings.
The result shows that the SAW seeps about $1.37\,\mathrm{mm}$ into the Bragg mirrors on both sides, namely, $l=0.36 + 2\times 1.37 = 3.1\,\mathrm{mm}$.
For the SAW resonator having the parameters listed in Table \ref{table:SAW_parameters}, we have $V=5.0\times10^{-11}\,\mathrm{m^3}$.
With the effective mode volume $V$, the zero-point-fluctuation $U_0$ can then be read as
\begin{equation}
    U_0 = \sqrt{\frac{\hbar}{2\rho V \omega_{\mathrm{SAW}}}} =6.5 \times 10^{-19}\,\mathrm{m},
\end{equation}
where $\rho=4.65\,\mathrm{g/cm^3}$ is the mass density of the $\mathrm{LiNbO_3}$.

On the other hand, the number of phonons $N$ can be written as
\begin{equation}
    N=\frac{4\kappa_{e1}}{(\kappa_{e1}+\kappa_{e2}+\gamma)^2}\frac{P_{\mathrm{rf}}}{\hbar \omega_{\mathrm{SAW}}}.
\end{equation}
From the fitting result described in Appendix~\ref{SM:S21} and input rf power $P_{\mathrm{rf}}=-11\,\mathrm{dBm}$, we have $N\sim 1.5\times10^{14}$.

Substituting the values of $U_0$ and $N$ into Eq.~(\ref{eq:uy}), we have
 \begin{equation}
     \tilde{u}_{y}(x,z)\sim \left(8.0 \times 10^{-12}\,\mathrm{m}\right) \times \sin(kz).
 \end{equation}
Thus, from Eq.~(\ref{eq:Tanduy}), the slope $\theta_{n}$ can be written as
\begin{equation}
    \theta_n(x,z) \sim \left(8.0 \times 10^{-12}\,\mathrm{m}\right)  \times\,k \cos(kz).
\end{equation}
Consequently, the maximum slope amplitude $|{\theta}_{n}|$ evaluated from the effective mode volume is as
\begin{equation}
   |{\theta}_{n}| \sim 1.3\times10^{-6}\,\mathrm{rad}.
\end{equation}
Considering that Figs.~2(b) and 2(c) are measured at the point with the largest slope variation, it is reasonable to compare the $|{\theta}_{n}|$ evaluated here with the $|{\theta}_{n}|$ evaluated with the other two methods.

%%221129


\begin{thebibliography}{99}

\bibitem{LL7}
L.~D.~Landau and E.~M.~Lifshitz, \textit{Theory of Elasticity}, 3rd ed, (Butterworth-Heinenann, Oxford, England, 1986).

\bibitem{TB2017}
K.~S.~Thorne and R.~D.~Blandford, \textit{Modern Classical Physics}, (Princeton University Press, Princeton, NJ, 2017).

%%%%%%%%%%%%%%%%%%%%%%%%%%%%%%%%%%%%%
\bibitem{C1998}
C.~Campbell, \textit{Surface Acoustic Wave Devices for Mobile and Wireless Communications}, (Academic, New York, 1998).

\bibitem{BB2008}
F.~Z.~Bi and B.~P.~Barber, ``Bulk acoustic wave RF technology'', IEEE Microwave Magazine~\textbf{9}, 65-80 (2008).

\bibitem{LR2008}
K.~L\"{a}nge, B.~E.~Rapp, and M.~Rapp, ``Surface acoustic wave biosensors: a review'', Analytical and Bioanalytical Chemistry~\textbf{391}, 1509-1519 (2008).

\bibitem{RM2009}
M.~Rocha-Gaso,~C.~March-Iborra, A.~Montoya-Baides, and A.~Arnau-Vives, ``Surface generated acoustic wave biosensors for the detection of pathogens: A review'' Sensors~\textbf{9}, 5740-5769 (2009).

\bibitem{R2017}
C.~C.~W.~Ruppel, ``Acoustic wave filter technology--a review'', IEEE Transactions on Ultrasonics, Ferroelectrics, and Frequency Control~\textbf{64}, 1390-1400 (2017).

\bibitem{ZL2021}
S.~I.~Zida, Y.~D.~Lin, and Y.~L.~Khung, ``Current Trends on Surface Acoustic Wave Biosensors'', Advanced Materials Technologies~\textbf{6}, 2001018 (2021).

%%%%%%%%%%%%%%%%%%%%%%%%%%%%%%%%%%%%%%%%%%%%%%%%%%%%%%%%%%%%%%%%%%%%%%

\bibitem{D2019}
P.~Delsing~\textit{et al}., ``The 2019 surface acoustic waves roadmap'', Journal of Physics D~\textbf{52}, 353001 (2019).

\bibitem{SK2015}
M.~J.~A.~Schuetz, E.~M.~Kessler, G.~Giedke, L.~M.~K.~Vandersypen, M.~D.~Lukin, and J.~I.~Cirac, ``Universal Quantum Transducers Based on Surface Acoustic Waves'', Physical Review X~\textbf{5}, 031031 (2015).

\bibitem{MK2017}
R.~Manenti, A.~F.~Kockum, A.~Patterson, T.~Behrle, J.~Rahamim, G.~ Tancredi, F.~Nori, and P.~J.~Leek, ``Circuit quantum acoustodynamics with surface acoustic waves'', Nature Communications~\textbf{8}, 975 (2017).

\bibitem{MS2018}
B.~A.~Moores, L.~ R.~Sletten, J.~J.~Viennot, and K.~W.~Lehnert,  ``Cavity Quantum Acoustic Device in the Multimode Strong Coupling Regime'', Physical Review Letters~\textbf{120}, 227701 (2018).

\bibitem{SZ2018}
K.~J.~Satzinger, Y.~P.~Zhong, H-S.~Chang, G.~A.~Peairs, A.~Bienfait, M.-H.~Chou, A.~Y.~Cleland, C.~R.~Conner, and {\'E}.~Dumur, J.~Grebel, I.~Gutierrez, B.~H.~November, R.~G.~Povey, S.~J.~Whiteley, D.~D.~Awschalom, D.~I.~Schuster, and A.~N.~Cleland, ``Quantum control of surface acoustic-wave phonons'', Nature~\textbf{563}, 661-665 (2018).

\bibitem{NY2020}
A.~Noguchi, R.~Yamazaki, Y.~Tabuchi, and Y.~Nakamura, ``Single-photon quantum regime of artificial radiation pressure on a surface acoustic wave resonator'', Nature Communications~\textbf{11}, 1183 (2020).

%%%%%%%%%%%%%%%%%%%%%%%%%%%%%%%%%%%%%%%%%%%%%%%%%%%%%%%%%%
\bibitem{W2002}
R.~F.~Wiegert, ``Magnetoelastic surface acoustic wave attenuation and anisotropic magnetoresistance in Ni3Fe thin films''Journal of Applied Physics~\textbf{91}, 8231 (2002).

\bibitem{WD2011}
M.~Weiler, L.~Dreher, C.~Heeg, H.~Huebl, R.~Gross, M.~S.~Brandt, and S.~T.~B.~Goennenwein, ``Elastically driven ferromagnetic resonance in nickel thin films'', Phys. Rev. Lett.~\textbf{106}, 117601 (2011).

\bibitem{WH2012}
M.~Weiler, H.~Huebl, F.~S.~Goerg, F.~D.~Czeschka, R.~Gross, S.~T.~
B.~G{\"o}ennenwein, ``Spin Pumping with Coherent Elastic Waves'', Physical Review Letters~\textbf{108}, 176601 (2012).

%gowtham
\bibitem{GM2015}
P.~G.~Gowtham, T.~Moriyama, D.~C.~Ralph, and R.~A.~Buhrman, ``Traveling surface spin-wave resonance spectroscopy using surface acoustic waves'', Journal of Applied Physics~\textbf{118}, 233910 (2015).

\bibitem{BV2020}
D.~A.~Bozhko, V.~I.~Vasyuchka, A.~V.~Chumak, and A.~A.~Serga, ``Magnon-phonon interactions in magnon spintronics'', Low Temperature Physics~\textbf{46}, 383-399 (2020).
%%%%%%%%%%%%%%%%%%%%%%%%%%%%%%%%%%%%%%%%%%%%%%%%%%
\bibitem{MC2012}
V.~Miseikis, J.~E.~Cunningham, K.~Saeed, R.~O'Rorke, and A.~G.~Davies, ``Acoustically induced current flow in graphen'', Applied Physics Letters~\textbf{100}, 133105 (2012).

\bibitem{PB2015}
T.~Poole, L.~Bandhu, and G.~R.~Nash, ``Acoustoelectric photoresponse in graphen'', Applied Physics Letters~\textbf{106}, 133107 (2015).

\bibitem{PS2015}
E.~Preciado, F.~J.~R.~Sch\"{u}lein, A.~E.~Nguyen, D.~Barroso, M.~Isarraraz, G.~von~Son, I-H.~Lu, W.~Michailow, B.~M\''{o}ller, V.~Klee, J.~Mann, A.~Wixforth, L.~Bartels, an H.~J.~Krenner, ``Scalable fabrication of a hybrid field-effect and acousto-electric device by direct growth of monolayer MoS2/LiNO3'', Nature communications~\textbf{6}, 1 (2015).

%%%%%%%%%%%%%%%%%%%%%%%%%%%%%%%%%%%%%%%%%%%%%%%%%

\bibitem{M2013}
M.~Matsuo, J.~Ieda, K.~Harii, E.~Saitoh, and S.~Maekawa, ``Mechanical generation of spin current by spin-rotation coupling'', Physical Review B~\textbf{87}, 180402(R) (2013).

\bibitem{IMM2014}
J.~Ieda, M.~Matsuo, S.~Maekawa, ``Theory of mechanical spin current generation via spin-rotation coupling'', Solid~State~Communications~\textbf{198}, 52-58 (2014).

\bibitem{K2017}
D.~Kobayashi, T.~Yoshikawa, M.~Matsuo, R.~Iguchi, S.~Maekawa, E.~Saitoh, and Y.~Nozaki, ``Spin Current Generation Using a Surface Acoustic Wave Generated via Spin-Rotation Coupling'', Physical~Review~Letters~\textbf{119}, 077202 (2017).

\bibitem{K2020}
Y.~Kurimune, M.~Matsuo, and Y.~Nozaki, ``Observation of Gyromagnetic Spin Wave Resonance in NiFe Films'', Physical~Review~Letters~\textbf{124}, 217205 (2020).
%%%%%%%%%%%%%%%%%%%%%%%%%%%%%%%%%%%%%%%%%%%%%%%%%%%%%%%%%%
% angular momentum
%\bibitem{ZN2014}
%L.~Zhang and Q.~Niu, ``Angular Momentum of Phonons and the Einstein–de Haas Effect'', Physical Review Letters~\textbf{112}, 085503 (2014).


\bibitem{LR2018}
Y.~Long, J.~Ren, and H.~Chen, ``Intrinsic spin of elastic waves'', Proc. Natl. Acad. Sci. U.S.A \textbf{115}, 9951 (2018).


\bibitem{SK2021}
M.M.~Sonner, F.~Khosravi, L.~Janker, D.~Rudolph, G.~Koblmüller, Z.~Jacob, and H.~J.~Krenner, ``Ultrafast electron cycloids driven by the transverse spin of a surface acoustic wave'', Science Advanced~\textbf{7}, eabf7414 (2021).


\bibitem{B2022}
K.Y.~Bliokh, ``Elastic Spin and Orbital Angular Momenta'', Physical~Review~Letters~\textbf{129}, 204303 (2022).

%%%%%%%%%%%%%%%%%%%%%%%%%%%%%%%%%%%%%%%%%%%%%%%%%%%%%%%%%

\bibitem{K2000}
J.~V.~Knuuttila, P.~T.~Tikka, and M.~M.~Salomaa, ``Scanning Michelson interferometer for imaging surface acoustic wave fields'', Optics Letters~\textbf{25}, 613-615 (2000).

\bibitem{H2011}
K.~A.~Hashimoto, ``Laser probe based on a sagnac interferometer with fast mechanical scan for RF surface and bulk acoustic wave devices'', IEEE Transactions on Ultrasonics, Ferroelectrics, and Frequency Control~\textbf{58}, 187-194 (2011).

\bibitem{FS2019}
W.~Fu, Z.~Shen, Y.~Xu, C.-H.~Zou, R.~Cheng, X.~Han, and H.~X.~Tang, ``Phononic integrated circuitry and spin--orbit interaction of phonons'', Nature Communications~\textbf{10}, 2743 (2019).

\bibitem{K2005}
H.~Kamizuma, L.~Yang, T.~Omori, K.~Hashimoto, and M.~Yamaguchi, ``High-speed laser probing system for surface acoustic wave devices based on knife-edge method'', Japanese Journal of Applied Physics~\textbf{44}, 4535 (2005).

\bibitem{H1970}
R.~J.~Hallermeier and W.~G.~Mayer, ``Light Diffraction by Ultrasonic Surface Waves of Arbitrary Standing-Wave Ratio'', The Journal of the Acoustical Society of America~\textbf{47}, 1236-1240 (1970).

\bibitem{IN2022}
A.~Iwasaki, D.~Nishikawa, M.~Okano, S.~Tateno, K.~Yamanoi, Y.~Nozaki, and S.~Watanabe, ``Temporal-offset dual-comb vibrometer with picometer axial precision'', APL Photonics~\textbf{7}, 106101 (2022).
%%%%%%%%%%%%%%%%%%%%%%%%%%%%%%%%%%%%%%%%%%%
\bibitem{TH2021}
K.~Taga, R.~Hisatomi, Y.~Ohnuma, R.~Sasaki, T.~Ono, Y.~Nakamura, and K.~Usami, ``Optical polarimetric measurement of surface acoustic waves'', Applied Physics Letters~\textbf{119}, 181106 (2021).


\bibitem{RB2004}
D.~Rugar, R.~Budakian, H.~J.~Mamin, and B.~W.~Chui, ``Single spin detection by magnetic resonance force microscopy'', Nature~\textbf{430}, 329 (2004). %dan


\bibitem{WH2021}
T.~Westphal, H.~Hepach, J.~Pfaff, and M.~Aspelmeyer, ``Measurement of gravitational coupling between millimeter-sized masses'', Nature~\textbf{591}, 225 (2021).


\bibitem{BF2003}
S.~M.~Barnett, C.~Fabre, and A.~Ma\^{i}tre, ``Ultimate quantum limits for resolution of beam displacements'', The European Physical journal D~\textbf{22}, 513 (2003).

\bibitem{CD2010}%kura-ku
A.~A.~Clerk, M.~H..~Devoret, S.~M.~Girvin, F.~Marquardt, and R.~J.~Schoelkopf, ``Introduction to quantum noise, measurement, and amplification'', Review of Modern Physics~\textbf{82}, 1155 (2010).

\bibitem{HO2016}%hisatomi
R.~Hisatomi, A.~Osada, Y.~Tabuchi, T.~Ishikawa, A.~Noguchi, R.~Yamazaki, K.~Usami, and Y.~Nakamura, ``Bidirectional conversion between microwave and light via ferromagnetic magnons'', Physical Review B~\textbf{93}, 174427 (2016).

\bibitem{JC1972}
P.~B.~Johnson and R.~W.~Christy, ``Optical constants of the noble metals'', Physical Review B~\textbf{6}, 4370 (1972).




\end{thebibliography}
\end{document}